\documentclass[a4paper]{article}

\usepackage{spconf,amsmath,graphicx,amssymb,subfig,booktabs,mathtools,caption}
\usepackage{INTERSPEECH2021}
\usepackage{graphicx}
\usepackage{amsmath}
\usepackage{xcolor}
\usepackage{glossaries}
\usepackage{derivative}

\usepackage{eucal}
\usepackage{dsfont}
\usepackage{multirow}
\usepackage{amsfonts}
\usepackage{array, pgfplots}
\pgfplotsset{compat=1.16}

\newacronym{mfcc}{MFCC}{Mel-frequency cepstral coefficient}
\newacronym{vae}{VAEs}{Variational Autoencoder}
\newacronym{gle}{GLE}{Group Latent Embedding}
\newacronym{vqvae}{VQVAE}{Vector Quantization VAE}
\newacronym{acvae}{ACVAE}{Auxiliary Classifier VAE}
\newacronym{mcd}{MCD}{Mel-Cepstral Distortion}
\newacronym{dtw}{DTW}{Dynamic Time-Warping}
\newacronym{vc}{VC}{Voice Conversion}
\newacronym{gan}{GAN}{Generative Adversarial Network}
\newacronym{mos}{MOS}{Mean Opinion Scores}
\newacronym{elbo}{ELBO}{Evidence Lower Bound}

\title{Many-to-Many Voice Conversion based Feature Disentanglement using Variational Autoencoder}
\name{Manh Luong$^1$, Viet Anh Tran$^2$}
\address{
  $^1$Vinai Research \\
  $^2$Deezer Research and Development, Paris, France}
\email{v.manhlt3@vinai.io, vatran@deezer.com}

\begin{document}

\maketitle
\begin{abstract}
  Voice conversion is a challenging task which transforms the voice characteristics of a source speaker to a target speaker without changing linguistic content. Recently, there have been many works on many-to-many~\acrfull{vc} based on~\acrfull{vae} achieving good results, however, these methods lack the ability to disentangle speaker identity and linguistic content to achieve good performance on unseen speaker's scenarios. In this paper, we propose a new method based on feature disentanglement to tackle many-to-many voice conversion. The method has the capability to disentangle speaker identity and linguistic content from utterances, it can convert from many source speakers to many target speakers with a single autoencoder network. Moreover, it naturally deals with the unseen target speaker's scenarios. We perform both objective and subjective evaluations to show the competitive performance of our proposed method compared with other state-of-the-art models in terms of naturalness and target speaker similarity.
\end{abstract}

\section{Introduction}

Voice conversion task is a technique which is used to transform a speaker's identity from the source speaker to the target speaker without changing the linguistic content~\cite{VC_definition}. The identity of a speaker, which depends on the voice timbre of a speaker, can be extracted from \acrfull{mfcc} by using statistical approaches~\cite{statistical_method}. These extracted features are then switched between source speakers and target speakers to perform ~\acrshort{vc}. Although the traditional methods succeed in ~\acrshort{vc} task~\cite{MCD,ClassicalVC1}, they require numerous parallel training data that are expensive to collect and require a lot of effort for alignment in order to get good results.

Two recent veins of work to overcome these issues are based on ~\acrshort{vae} and ~\acrfull{gan} frameworks.~\acrshort{gan} based approaches utilized an adversarial training procedure to learn a mapping function that is able to map from the source domain to the target domain. For solving many-to-many \acrshort{vc}, the authors in~\cite{StarGAN-VC2, CCGAN, CycleGAN2, CycleVAE} employed a cycle constraint to preserve the linguistic content of converted speech. Another way to improve the quality of converted speech is to extract speaker-independent information from input utterances by using an auxiliary classifier~\cite{MultitargetVC, StarGAN-VC2}. Although these approaches succeeded in transforming the source speaker's prosody into the target speaker's prosody~\cite{F0-Autovc},~\acrshort{gan} based methods are still difficult to train and the discriminator of~\acrshort{gan} does not resemble human auditory perception~\cite{F0-Autovc}. Moreover, the quality of converted voices are downgraded as more speakers are trained simultaneously~\cite{StarGAN-VC2}, hence~\acrshort{gan}-based method is ill-suited for many to many~\acrshort{vc} system.

~\acrshort{vae}-based methods have been adopted in order to model a latent space which is useful for interpolating between source and target domain. In~\cite{CycleVAE, vae-basedVC1, HierachicalVAEs, ACVAE, GLEVQVC}, the authors tried to model the latent space where latent factors are independent of one another by an encoder. The encoder usually is well-designed to remove speaker identity from latent vector~\cite{vae-basedVC1}, and then using a decoder conditioned with the target speaker identity, an utterance is generated which is compatible with conditioned speaker identity. Cycle-VAE~\cite{CycleVAE} proposed a cycle flow based method of~\acrshort{vae} to improve the quality of converted speech, however, it only performs one-to-one conversion which is an expensive method for transforming one source speaker into many target speakers. Since prior~\acrshort{vae}-based works used one-hot vectors to represent speaker identity, it could be problematic for performing~\acrshort{vc} for an unseen target speaker. Recent works~\cite{Autovc, F0-Autovc, VQVC+,Disentangle-Seq-Autoencoder, yuan2021improving} use the autoencoder framework to disentangle input voices into two parts: speaker representation, and phonetic content. During the conversion process, the source speaker's representation and the target speaker's representation are swapped to transform the source speaker's prosody to the target speaker's prosody. AutoVC-based methods~\cite{Autovc, F0-Autovc} were capable of converting the source speaker's tone to an unseen target speaker. Those methods leverage a universal pretrained speaker embedding and carefully designed a bottleneck, however, it is challenging to choose the proper bottleneck.

Therefore, we propose a disentangled~\acrfull{vae} based approach, called Disentangled-VAE, to tackle many-to-many~\acrshort{vc} by disentangling speaker-dependent information and linguistic information from the source and target speech. Our proposed method is able to deal with unknown speakers~\acrshort{vc} task. Our method is based on the assumption that there are some common factors shared among utterances coming from the same speaker, and that the remaining factors represent linguistic information which is distinctive from utterance to utterance. Specifically, during training, we sample a pair of utterances from the same speaker to feed into the encoder network to model a shared speaker latent vector by using an average function and two distinctive linguistic latent vectors. Then, the shared speaker latent vector is concatenated with the proper linguistic latent vector to reconstruct input utterances. For the conversion process, we swap the latent vector of the source speaker and the latent vector of the target speaker to generate the transformed speech. In the experiment section, both objective and subjective evaluations are conducted in VCTK Corpus~\cite{VCTK} to assess the performance of the proposed approach. The results show that Disentangled-VAE outperforms baseline methods in terms of~\acrfull{mcd} and the quality of converted speeches.

\section{Preliminary}

\label{sec:format}
\subsection{Conventional VAEs-based Non-Parallel VC}
\begin{figure}[t]
    \centering
    \subfloat[Conventional VAEs based VC]{%
    \includegraphics[width=2.6cm]{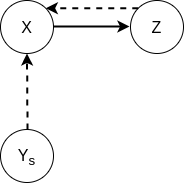}%
    \label{fig:1a}%
    }
    \subfloat[Disentanglement VAEs based VC]{%
    \includegraphics[width=2.6cm]{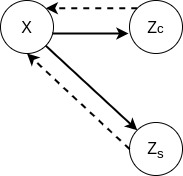}%
    \label{fig:1b}%
    }

    \caption{Directed graphical model of VAEs based VC. Solid arrows and dash arrows denote inference and generation process, respectively.}
    \label{fig:1}
\end{figure}

\acrshort{vae}~\cite{vae1,vae2} are a generative model which is used to model the probability density of data for the generating purpose. It includes two main components which are an encoder and a decoder parameterized by neural networks. The encoder models posterior distribution $q_{\phi}(z|x)$ of a latent variable $z$ given input data $x$, while the decoder approximates the conditional distribution $p_{\theta}(x|z)$ of data $x$ given a latent variable $z$. In~\cite{vae-basedVC1}, the latent code $\mathcal{z}$ has a strong assumption that it only contains the speaker-independent information,  such as phonetic content by carefully designing an encoder block. For performing the voice conversion shown in Figure. \ref{fig:1a}, a one-hot vector $y_s$, that represents the speaker identity concatenated with the latent vector $z$, is then fed into the decoder to convert source utterances to target utterances. Thus, the objective for \acrshort{vae}-based voice conversion is to maximize the marginal likelihood of the speech parameters $\theta$ given a speaker identity, $p_{\theta}(x|y_{s}) = \int p_{\theta}(x|z,y_{s})p_{\theta}(z)dz$, where $p_{\theta}(z)$ is the prior of the latent variables that usually is isotropic Gaussian distribution. Since the marginal likelihood is intractable to optimize, a lower bound is instead optimized for learning the encoder and decoder of \acrshort{vae} as follows:
\begin{equation}
\begin{aligned}
    \mathcal{L}(\phi,\theta,x,y_s) & = \mathbb{E}_{q_{\phi}(z|x)} \big[ \log{p}_{\theta}(x|z,y_s) \big] \\
    & -D_{KL}\big( q_{\phi}(z|x)||p_{\theta}(z)\big)
\end{aligned}
\end{equation}
where $D_{KL}$ denotes the Kullback-Leibler divergence between the approximate posterior distribution of latent vectors and the prior distribution. Both outputs of the encoder and the decoder are diagonal Gaussian distribution where mean and covariance are estimated by neural networks. 

\section{Proposed method}
The goal of disentanglement learning is to learn a controllable representation of data that is useful for interpolating between two real data samples. Therefore, this approach is well-fitted for style transfer, particularly for voice conversion tasks. Our proposed method, which is based on~\cite{WeaklySupervisedDW}, merely relies upon a weak assumption of data to learn disentangled latent space. We hypothesize that for a pair of utterances which come from a speaker,  there are some common factors representing speaker information and the remaining factors representing linguistic information as shown in Figure. ~\ref{fig:1b}. For learning disentangled latent space, the \acrshort{vae} framework is adopted to extract proper latent vector $z \in R^{k}$ given input utterance $x \in R^{d}$. The posterior distribution $p(z|x)$ is enforced by a weak assumption that there are two sub-latent spaces, each of them standing for a group of factors
\begin{equation}
    p(z_s|x_1) = p(z_s|x_2) \quad \forall z_s \in R^{k_1},
    \label{eq2}
\end{equation}
\begin{equation}
    p(z_c|x_1) \neq p(z_c|x_2) \quad \forall z_c \in R^{k_2},
    \label{eq3}
\end{equation}
, where $z_s$ and $z_c$ stand for the vector of speaker information and the vector of linguistic information, respectively. Both $R^{k_1} \subseteq R^{k}$ and $R^{k_2} \subseteq R^{k}$ are sub-spaces of latent space where $k_1+k_2=k$. In practice, $k_1$ is unknown but it can be chosen based on experiments, we will elaborate on the value of $k_1$ in the next section.

\begin{figure}[t]
    \centering
    \subfloat[Training process]{%
    \includegraphics[width=2.6cm]{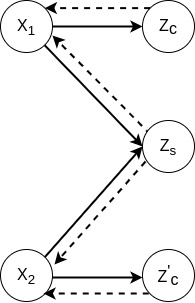}%
    \label{fig:2a}%
    }\qquad
    \subfloat[Conversion process]{%
    \includegraphics[width=2.6cm]{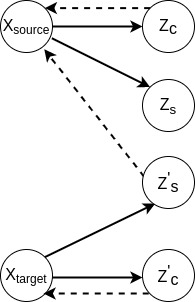}%
    \label{fig:2b}%
    }
    \caption{The training and conversion process for the proposed method. The solid and dash lines denote inference and generation procedure, respectively. }
    \label{fig:2}
\end{figure}

In order to force the constraint in Eq.\eqref{eq2}, the shared speaker vector $z_s$ is computed by a function $f(.)$ and then concatenated with linguistic vector $z_c$ to reconstruct the input $x$
\begin{equation}
\begin{aligned}
    z_s  \sim  q_{\phi}(z_s|x_1)& = \mathcal{N}\bigg( f(\mu_2(x_1),\mu_2(x_2)), f(\sigma_1(x_1),\sigma_2(x_2)) \bigg)
    \label{eq4}
\end{aligned}
\end{equation}
\begin{equation}
\begin{aligned}
    f(\mu{(x_1)}, \mu{(x_2)}) = \frac{\mu_1(x_1) + \mu_2(x_2)}{2}
\end{aligned}
\end{equation}
\begin{equation}
    z_c  \sim  q_{\phi}(z_c|x_1)
    \label{eq5}
\end{equation}
, where $q_{\phi}$ denotes an encoder network parameterized by $\phi$. The function $\mu(.)$ and $\sigma(.)$ estimate mean and variance of a posterior distribution, respectively. Function $f(.)$ is an average function proposed in~\cite{GVAE} which is successful in disentangling the common factor and the specific factor for images. The objective function is a variant of ~\acrfull{elbo} as follows
\begin{equation}
\begin{aligned}
    \mathcal{L}_{elbo}(\phi,\theta) & = \mathbb{E}_{(x_1,x_2)} \Big[\mathbb{E}_{q_{\phi}(z|x_1)}\log{p_{\theta}(x_1|z)} \\
    & + \mathbb{E}_{q_{\phi}(z|x_2)}\log{p_{\theta}(x_2|z)}  \\
    & - \beta D_{KL}(q_{\phi}(z|x_1)||p(z)) \\
    & - \beta D_{KL}(q_{\phi}(z|x_2)||p(z))  \Big] \\
\end{aligned}
\end{equation}
, where $p_{\theta}$ denotes a decoder network reparameterized by $\theta$, and latent vector $z$ is concatenated by two vector $z_s$ and $z_c$. Coefficient $\beta \geq 1$ which is analogous with $\beta -$VAE~\cite{betaVAE}.

We utilize a post net that is used in previous works~\cite{Autovc,tacotron2} to aid in creating fine details in reconstructed Mel-spectrograms. The post net comprises five convolutional layers with kernel size of $k=(5 \times 1)$. Its input is the decoder's output and the post net produces rough details of Mel-spectrogram. The first four layers have 512 channel dimensions, a batch norm and tanh activation are operated for those layers. The channel dimension of the last layer is 80 channels, to squeeze outputs to have an identical size with input. The output of post net is then added with the output of the decoder to reconstruct the final Mel-spectrogram.
\begin{equation}
    \hat{X} = \Tilde{X} + \Bar{X}
\end{equation}
, where $\Tilde{X}$ and $\Bar{X}$ are the output of the decoder and the post net respectively. $\hat{X}$ is the final reconstruction of Mel-spectrogram. Consequently, we have an additional reconstruction loss as follows:
\begin{equation}
    \mathcal{L}_{recons} = || X - \hat{X} ||
\end{equation}
, where $X$ is ground truth Mel-spectrogram. Finally, to train Disentangled-VAE, we use the total loss function as follows:
\begin{equation}
    \mathcal{L} = \mathcal{L}_{elbo} + \mathcal{L}_{recons}
\end{equation}

Figure.~\ref{fig:2a} depicts the training process of our proposed method. There are two utterances from the same speaker fed into the encoder to encode a common factor (speaker identity) $z_s$ and a specific factor (phonetic content) $z_c$. Subsequently, for the generation procedure, a speaker identity vector is concatenated with two content vectors to reconstruct two input utterances. For the conversion process shown in Figure.~\ref{fig:2b}, after inferring two speaker identity vectors of the source and the target utterances, the target speaker identity is concatenated with the content vector of the source utterance before feeding into the decoder to acquire the converted utterance.  
\section{Experimental settings}

\label{sec:experimental settings}
We use VCTK Corpus which includes 109 English speakers with a variety of accents for our experiment. The dataset is split into two parts a training set and a testing set, in which there are 105 speakers selected for training and the remaining 4 speakers are used for one-shot conversion testing. There are 20 utterances of training speakers reserved for testing seen speakers voice conversion. To evaluate seen speakers conversion performance, we randomly choose 4 speakers: p225(female), p226(male), p229(female), and p232(male). To measure one-shot speakers conversion performance, we measure converted utterance on the testing dataset, which includes 4 random speakers: p282(female), p286(male), p294(female), and p334(male). There are 4 experiments conducted as male-male, male-female, female-female, and female-male for both objective and subjective evaluation.

All utterances are sampled at 16k Hz. We opt to extract STFT features with a Hamming window of size 1024 samples and a hop length of size 256 samples. Next, Mel-spectrograms are converted to 80 bins from extracted STFTs and take logarithm. The Mel-spectrograms are then normalized with a range from 0 to 1 for stability training. At every training iteration, a pair of Mel-spectrogram segments of size $64 \times 80$ are randomly extracted from two utterances coming from a speaker as input fed into the neural network. We acquire a pair of utterances by sampling from the uniform distribution of the speaker's utterances. For reverting Mel-spectrograms to waveform signals, we apply the Wavenet vocoder~\cite{Wavenet} to generate a proper waveform from a given Mel-spectrogram. 

Table~\ref{tab:1} details the architecture of the proposed model. The proposed architecture technically has two main components: an encoder and a decoder. For the encoder, it consists of three 1D convolutional layers, subsequently, there is a BiLSTM layer and three fully connected layers. the encoder outputs two latent vectors which represent the speaker identity $Z_s \in \mathbb{R}^8$ and the phonetic content $Z_c \in \mathbb{R}^{56}$. The dimension of speaker identity $k_1=8$, which is chosen according to empirical experiment. For the decoder, it comprises of three fully connected layers, two LSTM layers, three 1D convolutional layers. At the last layer of the decoder, the decoder outputs a reconstructed Mel-spectrogram. We train Disentangled VAE on an Nvidia V100 GPU with a batch size of 8 samples for about 1M training steps. We use Adam optimizer with learning rate $lr=1e-4$   for the whole training phase. The pretrained model and audio samples can be found in our github\footnote{https://v-manhlt3.github.io/disentangled-VAE/}
\begin{table}[ht!]
\small
\caption{The architecture of the proposed model. The parameters of the 1D convolution layer are denoted as kernel size, stride, output channel. The parameters of the LSTM layer are input size, hidden size, and the number of hidden layers.}
\label{tab:1}
\begin{tabular}{|l|l|l|}
\hline
\multicolumn{1}{|c|}{Block}                    & \multicolumn{2}{c|}{Layer}                                                     \\ \hline
\multicolumn{1}{|c|}{\multirow{5}{*}{Encoder}} & Input layer              & 64 $\times$ 80    \\ \cline{2-3} 
\multicolumn{1}{|c|}{}                         & 1D Conv layer $\times 3$   & (5, 2, 512)        \\ \cline{2-3} 
\multicolumn{1}{|c|}{}                         & BiLSTM layer             & (512, 64, 2)       \\ \cline{2-3} 
\multicolumn{1}{|c|}{}                         & FC layer                 & (2048, 256)       \\ \cline{2-3} 
\multicolumn{1}{|c|}{}                         & FC layer (phonetic content)  & (256, 56)         \\ \cline{2-3}
\multicolumn{1}{|c|}{}                         & FC layer  (speaker)      & (256, 8)           \\ \hline
\multirow{6}{*}{Decoder}                       & FC layer                 & (32, 256)        \\ \cline{2-3} 
                                               & FC layer                 & (256, 2048)       \\ \cline{2-3} 
                                               & LSTM layer               & (64, 512, 1)       \\ \cline{2-3} 
                                               & 1D Conv layer $\times 3$ & (5, 2, 512)    \\ \cline{2-3} 
                                               & LSTM layer               & (512, 1024, 2)     \\ \cline{2-3} 
                                               & FC layer                 & (1024, 80)        \\ \hline
\end{tabular}
\end{table}
\begin{figure}[ht]
    \centering
    \includegraphics[width=0.45\textwidth]{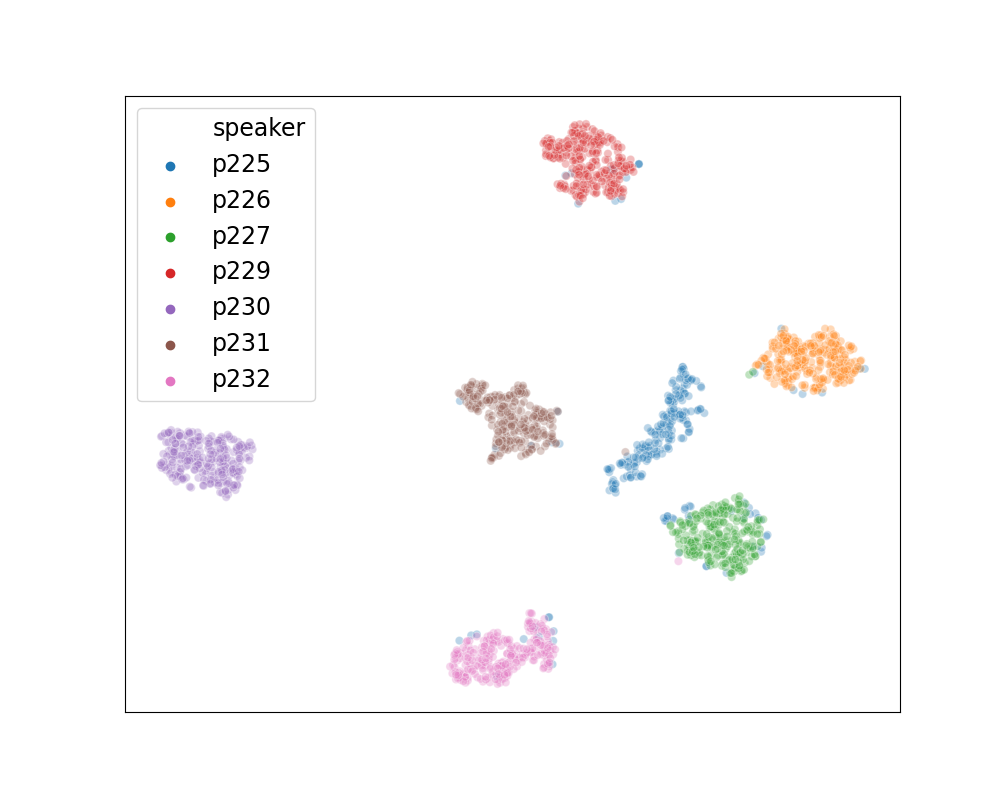}
    \caption{tSNE Visualization of speaker embedding space.}
    \label{fig:3}
\end{figure}
\subsection{Results}
\label{sec:results}
We evaluate the performance of the proposed method through both objective and subjective evaluation compared with three baseline models: Autovc~\cite{Autovc}, VQVC+~\cite{VQVC+}, and CycleVAE~\cite{CycleVAE} that is the baseline system for VCC2020~\cite{vcc2020}. Both Autovc and VQVC+ are autoencoder based approaches for VC, also, they try to disentangle speaker-independent information from input utterances by learning a well-designed encoder. Since CycleVAE is only capable of converting seen speakers, thus we evaluate its performance on seen speaker conversion. The objective metric used to measure the performance of converted speech is~\acrshort{mcd} as described in~\cite{MCD}. It estimates the discrepancy between two aligned utterances. We therefore align both target ground truth and converted utterance by using~\acrfull{dtw}, there are 12 first utterances of all speakers evaluated since they have the same contents. Regarding subjective evaluation, we conducted two sorts of tests which are~\acrfull{mos} and ABX testing. For~\acrshort{mos}, there were 15 participants who were asked to assess the quality of 20 utterances per system by selecting scores from 1-bad to 5-perfect. For ABX testing, those participants are asked to rate 25 pairs of conversion utterances for each experimental group. During subjective experiments, participants wear the same headphones for all system evaluation to achieve a fair comparison among all systems.
\begin{table}[ht!]
\small
\centering
\caption{Comparison of ~\acrshort{mcd} and MOS for seen speakers among three methods and CycleVAE that is baseline system in VCC2020.}
\label{tab:2}
\begin{tabular}{c | c | c | c | c | c | c } 
\toprule
\multirow{2}{*}{Method} & \multirow{2}{*}{MCD)} &\multicolumn{5}{c}{MOS}  \\ \cline{3-7} 
                        &                      & M2M  & M2F  & F2M  & F2F & Avg \\ \hline
VQVC+                   &        9.63         &    3.09  & 2.97     &  2.88    &  2.93 & 2.97  \\ \hline
AutoVC                  &        8.82         &    3.11  & \textbf{3.35}    &   2.97   &  3.29 & 3.18   \\ \hline
CycleVAE          &        9.77         &    2.71  & 2.68    &   2.81   &  2.87 & 2.76   \\ \hline
Proposed                &     \textbf{7.51}    &   \textbf{3.87}   & 3.21 & \textbf{3.25}& \textbf{3.59} & \textbf{3.48}   \\ \bottomrule
\end{tabular}
\end{table}
\begin{table}[ht!]
\small
\centering
\caption{Comparison of ~\acrshort{mcd} and MOS for unseen speakers among baseline systems. Since CycleVAE is incapable of performing one-shot voice conversion, we do not report its result in this table.}
\label{tab:3}
\begin{tabular}{c | c | c | c | c | c | c} 
\toprule
\multirow{2}{*}{Method} & \multirow{2}{*}{MCD)} &\multicolumn{5}{c}{MOS}  \\ \cline{3-7} 
                        &                      & M2M  & M2F  & F2M  & F2F & Avg \\ \hline
VQVC+                   &       10.01     &   \textbf{3.18}   & \textbf{3.06}     &  2.87    &  2.91 & 3.01   \\ \hline
AutoVC                  &       10.48      &    2.76  & 2.82 &  2.63    &  2.85 & 2.76   \\ \hline
CycleVAE          &    -    &   -  & -    &   -   &  - & -   \\ \hline
Proposed                &       \textbf{9.72}  &   3.12   &  3.03     &  \textbf{3.23}     & \textbf{2.96} & \textbf{3.09}   \\ \bottomrule
\end{tabular}
\end{table}


Table. 2 presents the experiment result for seen speaker conversion in terms of~\acrshort{mcd} and~\acrshort{mos}. As shown in the table, our proposed system outperforms baseline systems on MCD which is 7.51 dB. Regarding MOS, our system achieves a significant improvement on speech  quality  in  most cases for seen speaker conversion, but AutoVC achieves marginally better conversion performance in male-to-female circumstances. Table. 3 shows the comparison with regard to~\acrshort{mcd} and~\acrshort{mos} for unseen speaker conversion conditions. Disentangled VAE marginally surpasses baseline systems in terms of~\acrshort{mcd} which is 9.72 dB. In terms of MOS, our model generates converted voice with high quality for two cases: female-to-male and female-to-female. For other cases, VQVC+ slightly outperforms our model, however, on average the proposed model produces converted speech with better quality than baseline systems regarding subjective evaluation.

\begin{figure}[htp]
\label{fig:4}
\subfloat[Average score(\%) of conversion speech on naturalness.]{%
  \includegraphics[clip,width=\columnwidth]{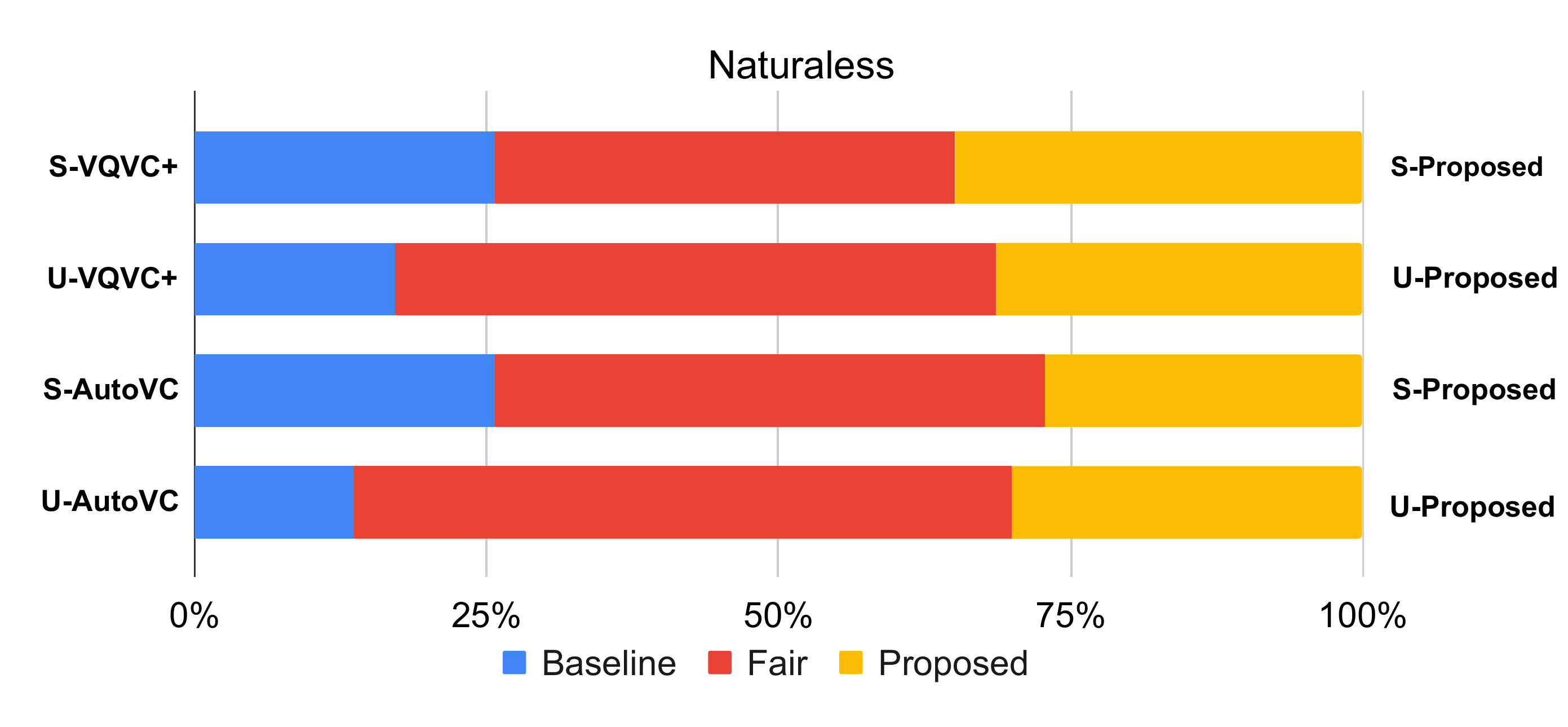}%
  \label{fig:4a}
}

\subfloat[Average score(\%) of conversion speech on similarity]{%
  \includegraphics[clip,width=\columnwidth]{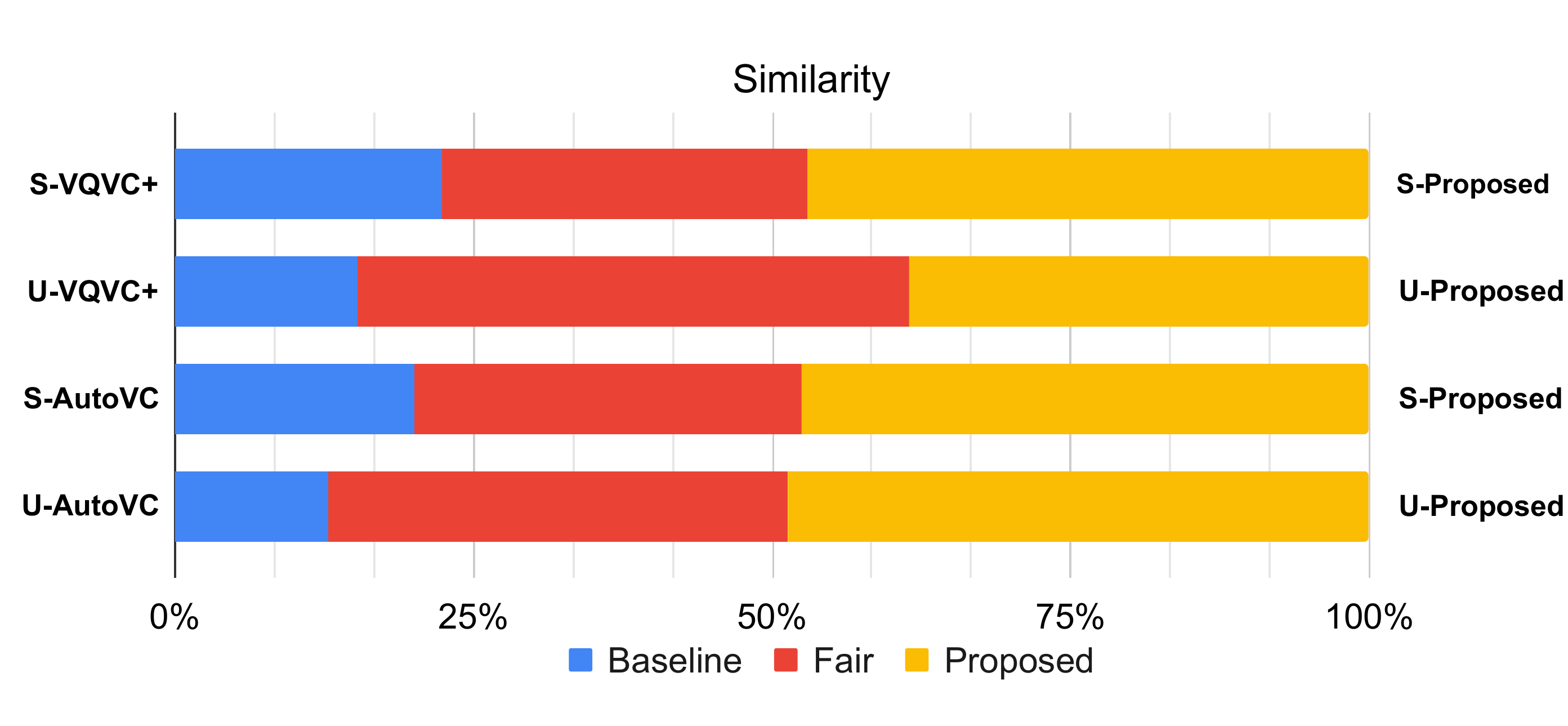}%
  \label{fig:4b}
}
\caption{ABX testing experimental results on Naturalness and Similarity for pairwise system comparison. "S-" and "U-" denote seen speakers and unseen speakers during training. }
\end{figure}
Fig.~\ref{fig:3} illustrates speaker embedding visualized by tSNE method, there are 30 utterances sampled for every speaker to calculate the speaker representation. According to the empirical results, we found that a chunk of 2 seconds is adequate to extract the speaker representation. As shown in Fig.~\ref{fig:3}, speaker embeddings are separable for different speakers. In contrast, the speaker embeddings of utterances of the same speaker are close to each other. As a result, our method is able to extract speaker-dependence information by using the encoder network. Fig.~\ref{fig:4a} and Fig.~\ref{fig:4b} demonstrate the ABX testing between the Disentangled-VAE system and baseline systems. Because Cycle-VAE is incapable of performing one shot voice conversion, we do not experiment with ABX testing for that system. If you are interested in listening to converted utterances of baseline systems and our proposed method, please visit our demo page. As shown in these figures, our system significantly outperforms AutoVC and VQVC+ on similarity score, in particular, its conversion utterances for unseen speakers are much better than the baselines in terms of naturalness and similarity to target speakers.

\section{Conclusions}
In this paper, we propose a feature disentanglement approach using~\acrshort{vae} framework for voice conversion. By leveraging the hypothesis that there are some shared factors among speech from the same speaker. As a result, our method is able to disentangle speaker identity and phonetic content from input utterances. Moreover, thanks to its capability of disentanglement, Disentangled-VAE can transform voice from a source speaker to an unknown target speaker. The empirical experiment shows that Disentangled-VAE acquires competitive performance compared with state-of-the-art solutions in terms of both naturalness of the converted speeches and its similarity with the target speaker's voice. In addition, while disentangled speaker identities of utterances look separable for different speakers, on the contrary, they are close to one another. Therefore, we believe that disentangled approaches are promising for not only voice conversion but also speaker verification and speaker diarization.

\bibliographystyle{IEEEtran}

\bibliography{refs}

\end{document}